\documentclass[letterpaper, 10 pt, conference]{ieeeconf}  %
\pdfoutput=1 %

\IEEEoverridecommandlockouts                              %

\usepackage{graphics} %
\usepackage{epsfig} %
\usepackage{amsmath} %
\usepackage{amssymb}  %
\usepackage{leftidx}
\usepackage{multirow}
\usepackage[table,xcdraw]{xcolor}
\usepackage{subcaption}
\usepackage{algpseudocode}
\usepackage{algorithm}

\usepackage{xspace}
\makeatletter
\DeclareRobustCommand\onedot{\futurelet\@let@token\@onedot}
\def\@onedot{\ifx\@let@token.\else.\null\fi\xspace}

\def\eg{\emph{e.g}\onedot} 
\def\ie{\emph{i.e}\onedot}

\makeatother

\usepackage{amsmath}

\newcommand*{\tran}{^{\mkern-1.5mu\mathsf{T}}}
\newcommand*{\minustran}{^{\mkern-1.5mu\mathsf{-T}}}
\newcommand{\norm}[1]{\left\lVert#1\right\rVert}
\newcommand{\crossmat}[1]{\lfloor#1\rfloor_\times}

\newcommand{\Exp}{\mathrm{Exp}}
\newcommand{\Log}{\mathrm{Log}}
\newcommand{\mbf}[1]{\mathbf{#1}}

\makeatletter
\let\NAT@parse\undefined
\makeatother
\usepackage[pagebackref=true,bookmarks=false,hidelinks]{hyperref}

\title{\LARGE \bf
A Quick Guide for the Iterated Extended Kalman Filter on Manifolds
}

\author{Jianzhu Huai$^{1}$, %
	Xiang Gao$^{2}$
	\thanks{$^{1}$Jianzhu Huai is with the State Key Lab of Surveying, Mapping, and Remote Sensing, Wuhan University}%
	\thanks{$^{2}$Xiang Gao was with the Technical University of Munich.}
}

\begin{document}

\maketitle
\thispagestyle{empty}
\pagestyle{empty}

\begin{abstract}
The extended Kalman filter (EKF) is a common state estimation method for discrete nonlinear systems.
It recursively executes the propagation step as time goes by and the update step when a set of measurements arrives.
In the update step, the EKF linearizes the measurement function only once.
In contrast, the iterated EKF (IEKF) refines the state in the update step by iteratively solving a least squares problem.
The IEKF has been extended to work with state variables on manifolds which have differentiable $\boxplus$ and $\boxminus$ operators, including Lie groups.
However, existing descriptions are often long, deep, and even with errors.
This note provides a quick reference for the IEKF on manifolds, using freshman-level matrix calculus.
Besides the bare-bone equations, we highlight the key steps in deriving them.
\end{abstract}

\section{Problem and Assumptions}
The problem is to estimate the state vector $\mbf x$ and its covariance $\mbf P$ given its dynamic equation $\mbf f()$ 
and the external observations $\mbf z$.
We assume that the state vector $\mbf x$ is an element of a differentiable manifold with a boxplus operator $\boxplus$ and a boxminus operator $\boxminus$, defined as
\begin{align}
	\mbf x \boxplus \boldsymbol{\delta} = \mbf y, \\
	\mbf y \boxminus \mathbf{x} = \boldsymbol{\delta}.
\end{align}
where $\mbf y$ is another manifold element close to $\mbf x$ with a distance vector $\boldsymbol{\delta}$.
For an Euclidean space, $\boxplus$ is the common plus(+) and $\boxminus$ is the common minus (-).
For elements of SO3, example definitions of $\boxplus$ and $\boxminus$ are given in \eqref{eq:fastlio-so3} and \eqref{eq:huai-so3}.
In the state estimation problem, we usually denote the estimate of $\mbf x$ by $\widehat{\mbf x}$,
which is usually defined relative to $\mbf x$ by
\begin{align}
\widehat{\mbf x} \boxplus \boldsymbol{\delta} = \mbf x, \\
\mbf x \boxminus \widehat{\mathbf{x}} = \boldsymbol{\delta}.	
\end{align}
An alternative definition is $\mbf x \boxplus \boldsymbol{\delta} = \widehat{\mbf x}$, \eg, in \cite{jekeli_inertial_2001}.
This definition leads to a less-intuitive Kalman update step and is uncommonly used.

For the extended Kalman filter (EKF), we use the dynamic equation discretized at time steps $t_1, \cdots, t_k$.
Time steps are usually chosen according to the stamp of the measurement vector $\mbf z$, \eg, $\mbf z_k = \mbf z(t_k)$ is acquired at time $t_k$.
Let's consider one EKF time step from $t_{k-1}$ until $t_k$, which involves propagation with the discrete dynamic equation and update with the measurement $\mbf z_k$.
The discrete dynamic equation from $t_{k-1}$ to $t_k$ is given by
\begin{equation}
\mbf x_k = \mbf f(\mbf x_{k-1}, \mbf u_{k-1}, \mbf w_{k-1}),
\end{equation}
where $\mbf w_{k-1}$ is the discrete noise at $t_{k-1}$ with a Gaussian distribution $N(\mbf 0, \mbf Q_{k-1})$.
The observation equation at $t_k$ is
\begin{equation}
	\mbf z_k = \mbf h(\mbf x_k) + \mbf n_k,
\end{equation}
where $\mbf n_{k}$ is the discrete noise at $t_{k}$ with a Gaussian distribution $N(\mbf 0, \mbf R_k)$.
For simplicity, we assume that the observation noise is additive and $\mbf z_k$ is in a vector space.
Indeed, many state estimation problems meet this assumption.

\section{Iterated Extended Kalman Filter}
Now we describe the estimation process of the iterated EKF (IEKF).
Denote the a-posteriori state vector estimate at $t_{k-1}$ by $\mathbf{x}_{k-1}^+$ and its covariance by $\mbf P_{k-1}^+$.
The propagation step propagates the state estimate to $t_k$ by
\begin{align}
	\mbf x_{k}^- &= \mbf f(\mbf x_{k-1}^+, \mbf u_{k-1}, \mbf 0) \label{eq:prop}, \\
	\mbf P_k^- &= \mbf F_{k-1} \mbf P_{k-1}^+ \mbf F_{k-1}\tran + \mbf G_{k-1} \mbf Q_{k-1} \mbf G_{k-1}\tran \label{eq:prop-cov},
\end{align}
where the propagation Jacobians are
\begin{align}
\mbf F_{k-1} &= \lim_{\boldsymbol{\epsilon}\rightarrow\mbf 0}
\frac{\mbf f(\mbf x_{k-1}^+ \boxplus \boldsymbol{\epsilon}, \mbf u, \mbf w) \boxminus \mbf f(\mbf x_{k-1}^+, \mbf u, \mbf w)}
{\boldsymbol{\epsilon}} \\
\mbf G_{k-1} &= 
\frac{\partial \mbf f(\mbf x_{k-1}^+, \mbf u, \mbf w)}
{\partial \mbf w}
\Big\vert_{\mbf w = \mbf 0}
\end{align}

With the measurement $\mbf z_k$, the EKF solves for the state vector $\mbf x_{k}^+$ that minimizes the weighted square sum of the deviation from the predicted state vector $\mbf x_k^-$ and 
the innovation term $\mbf r_k = \mbf z_k - \mbf h(\mbf x_k^+)$, \ie,
\begin{equation}
\begin{split}
\min_{\mbf x_k^+}
&(\mbf x_k^+ \boxminus \mbf x_k^-)\tran {\mbf P_k^-}^{-1} (\mbf x_k^+ \boxminus \mbf x_k^-) + 
\\ 
&(\mbf z_k - \mbf h(\mbf x_k^+))\tran \mbf R_k^{-1} (\mbf z_k - \mbf h(\mbf x_k^+)).
\end{split}
\end{equation}

In contrast, the IEKF iteratively updates the state vector starting from 
$\mbf x_{k, 0}^+ = \mbf x_k^-$.
In iteration $j$, it solves for an increment $\boldsymbol{\delta}_{k, j}$ to the current state vector estimate $\mbf x_{k, j}^+$ that minimizes the weighted square sum of
the deviation from the prediction $\mbf x_k^-$ and the innovation $\mbf r_{k,j} = \mbf z_k - \mbf h(\mbf x_{k,j}^+)$,
\begin{equation}
\begin{split}
\min_{\boldsymbol{\delta}_{k, j}}
&\norm{\mbf x_{k, j}^+ \boxminus \mbf x_k^- + \mbf J_{k,j} \boldsymbol{\delta}_{k, j}}
_{{\mbf P_k^-}^{-1}} + \\
&\norm{\mbf z_k - \mbf h(\mbf x_{k,j}^+) - \mbf H_{k, j} \boldsymbol{\delta}_{k,j}}_
{\mbf R_k^{-1}}
\end{split}\label{eq:min-iekf}
\end{equation}
where the Jacobian matrices are defined as
\begin{align}
\mbf H_{k,j} &= \lim_{\boldsymbol{\epsilon}\rightarrow \mbf 0}
\frac{\mbf h(\mbf x_{k, j}^+ \boxplus \boldsymbol{\epsilon}) - \mbf h(\mbf x_{k, j}^+)}{\boldsymbol{\epsilon}} \\
\mbf J_{k,j} &= \lim_{\boldsymbol{\epsilon}\rightarrow\mbf 0}
\frac{(\mbf x_{k,j}^+ \boxplus \boldsymbol{\epsilon} \boxminus \mbf x_k^-) - (\mbf x_{k,j}^+ \boxminus \mbf x_k^-)}{\boldsymbol{\epsilon}}
\end{align}
$\mbf J_{k,j}$ is a square identity matrix if $\mbf x$ is in a vector space. 
Otherwise, it can be well approximated by the square identity matrix.
Setting the derivative of \eqref{eq:min-iekf} to zero, we can get the solution for $\boldsymbol{\delta}_{k,j}$, which are used for the iterative update\footnote{Thanks to Yarong Luo from WHU for checking the update equations.},
\begin{align}
\mbf S_{k,j} &= \mbf H_{k,j} \mbf L_{k,j} \mbf P_{k}^- \mbf L_{k,j}\tran \mbf H_{k,j}\tran + \mbf R_k \label{eq:s},\\
\mbf K_{k,j} &= \mbf L_{k,j} \mbf P_{k}^- \mbf L_{k,j}\tran \mbf H_{k,j}\tran \mbf S_{k,j}^{-1} \label{eq:k}, \\
\begin{split}
\boldsymbol{\delta}_{k,j} &= \mbf K_{k,j}\big[\mbf H_{k,j} \mbf L_{k,j}(\mbf x_{k,j}^+ \boxminus \mbf x_k^-) + \\ &\mbf z_k - \mbf h(\mbf x_{k,j}^+)\big]
-\mbf L_{k,j}(\mbf x_{k,j}^+ \boxminus \mbf x_k^-),
\end{split}\\
\mbf x_{k,j+1}^+ &= \mbf x_{k,j}^+ \boxplus \boldsymbol{\delta}_{k,j}.
\label{eq:state-update}
\end{align}
where $\mbf L_{k,j} = \mbf J_{k,j}^{-1}$.
The covariance will be updated once at the end of all iterations,
\begin{equation}
\label{eq:cov-update}
\mbf P_{k}^+ = (\mbf I - \mbf K_{k,n} \mbf H_{k,n}) \mbf L_{k,n} \mbf P_k^- \mbf L_{k,j}\tran
\end{equation}

In summary, the IEKF algorithm is given in Algorithm \ref{alg:iekf}.
\begin{algorithm}
\caption{The iterated extended Kalman filter algorithm}\label{alg:iekf}
\begin{algorithmic}
	\State \textbf{Input} at time $t_{k-1}$: $\mbf x_{k-1}^+$, $\mbf P_{k-1}^+$, $\mathbf u_{k-1}$, $\mbf z_k$, max iterations $n$, termination threshold $\epsilon$
    \State \textbf{Output} at time $t_{k}$: $\mbf x_{k}^+$, $\mbf P_{k}^+$
	\State Propagate state and covariance with \eqref{eq:prop}\eqref{eq:prop-cov} to $\mbf x_{k}^-$, $\mbf P_{k}^-$
	\State $j \gets 0$
    \State	$\mbf x_{k, 0}^+ \gets \mbf x_{k}^-$
	\While{$j \neq n$}
		\State update the state vector estimate with \eqref{eq:s}-\eqref{eq:state-update}
		\State $j \gets j + 1$
		\If{$\norm{\boldsymbol{\delta}_{k,j}} < \epsilon$}
		\State break
		\EndIf
	\EndWhile
	\State $\mbf x_{k}^+ =  \mbf x_{k, n}^+$
    \State update the covariance with \eqref{eq:cov-update}
\end{algorithmic}
\end{algorithm}

When the dimension of observations is high, the inversion in \eqref{eq:k} can be intensive. There are two approaches to deal with this issue.
One is to perform QR decomposition of $\mbf H_{k,j}$ as done in MSCKF \cite[(4.61)]{huai_collaborative_2017},
\begin{align}
\mbf H_{k,j} &= \begin{bmatrix}
	\mbf Q_1 & \mbf Q_2
\end{bmatrix}
\begin{bmatrix}
	\mbf T_H \\ \mbf 0
\end{bmatrix}, \\
\underbrace{\mbf Q_1\tran (\mbf z - \mbf h(\mbf x_{k,j}^+))}_{\mbf r_q} &= \mbf T_H \boldsymbol{\delta}_{k,j} + \underbrace{\mbf Q_1\tran \mbf n_k}_{\mbf n_q},
\end{align}
where $\mbf n_q$ is a Gaussian noise, $N(\mbf 0, \mbf R_{q} = \mbf Q_1\tran \mbf R_k \mbf Q_1)$.
Then, substituting $\mbf r_q$, $\mbf T_H$, and $\mbf R_q$ for $\mbf r_{k,j}$, $\mbf H_{k,j}$, and $\mbf R_k$, the IEKF update proceeds as in \eqref{eq:s}-\eqref{eq:cov-update}.
The other approach is to rewrite $\mbf K$ as done in FAST-LIO \cite{xu_fast-lio_2021}, \ie,
\begin{equation}
\mbf K_{k,j} = [\mbf H_{k,j}\tran \mbf R_k^{-1} \mbf H_{k,j} + (\mbf L_{k,j}\mbf P_k^{-} \mbf L_{k,j}\tran)^{-1}]^{-1} \mbf H_{k,j}\tran \mbf R_k^{-1}.
\end{equation}
The expression is equivalent to \eqref{eq:k}, but needs an extra covariance inversion.

\section{Example Manifold SO3}
As an example, we compute $\mbf F_{k-1}$ and $\mbf J_{k,j}$ when $\mbf x$ is an SO3 element.
Denote the orientation of $\{B\}$ frame relative to the $\{W\}$ frame by $\mbf{R}_{WB}$, and its estimate by $\widehat{\mbf{R}}_{WB}$. 
If we define 
$\boxplus$ and $\boxminus$ like
\begin{align}
\label{eq:fastlio-so3}
\widehat{\mbf{R}}_{WB} \boxplus \delta\theta_{WB} &= \widehat{\mbf{R}}_{WB} \Exp(\delta\theta_{WB}) = \mbf{R}_{WB}, \\
\mbf{R}_{WB} \boxminus \widehat{\mbf{R}}_{WB} &= \Log(\widehat{\mbf{R}}_{WB}\tran \mbf{R}_{WB}) =  \delta\theta_{WB},
\end{align}
as done in FAST-LIO \cite{xu_fast-lio_2021}, then
$\mbf F_{k-1}$ and $\mbf J_{k,j}$ are
\begin{align}
\begin{split}
\mbf F_{k-1} &= \lim_{\boldsymbol{\epsilon}\rightarrow\mbf 0}
\frac{1}{\boldsymbol{\epsilon}}
\Log\Big[ (\mbf R_{WB_{k-1}}\Exp(\int_{t_{k-1}}^{t_k}\boldsymbol{\omega}_{WB}^Bdt))\tran \cdot \\
&\mbf R_{WB_{k-1}} \Exp(\boldsymbol{\epsilon}) \Exp(\int_{t_{k-1}}^{t_k}\boldsymbol{\omega}_{WB}^Bdt)\Big] \\
&= \Exp(-\int_{t_{k-1}}^{t_k}\boldsymbol{\omega}_{WB}^Bdt),
\end{split} \\ 
\begin{split}
\mbf J_{k,j} &= \lim_{\boldsymbol{\epsilon}\rightarrow\mbf 0} \frac{\Log({\mbf x_k^-}\tran \mbf x_{k,j}^+ \Exp(\boldsymbol{\epsilon})) - \Log({\mbf x_k^-}\tran \mbf x_{k,j}^+)}{\boldsymbol{\epsilon}} \\
&=\mbf J_r^{-1}(\delta \phi),
\end{split} \\
\delta \phi_j &= \mbf x_{k, j}^+ \boxminus \mbf x_k^-,
\end{align}
where we use the angular velocity $\boldsymbol{\omega}_{WB}^B$ relationship 
$\mbf R_{WB_{k}} = \mbf R_{WB_{k-1}} \Exp(\int_{t_{k-1}}^{t_k}\boldsymbol{\omega}_{WB}^Bdt)$,
and $\mbf x$ denotes the rotation matrix $\mbf R_{WB}$ for clarity.
The above results agree with FAST-LIO equations.

Otherwise, if we define $\boxplus$ and $\boxminus$ like
\begin{align}
	\label{eq:huai-so3}
	\widehat{\mbf{R}}_{WB} \boxplus \delta\theta_{WB} &= \Exp(\delta\theta_{WB}) \widehat{\mbf{R}}_{WB} = \mbf{R}_{WB} \\
	\mbf{R}_{WB} \boxminus \widehat{\mbf{R}}_{WB} &= \Log(\mbf{R}_{WB} \widehat{\mbf{R}}_{WB}\tran)
\end{align} as in \cite{huai_observability_2022},
then $\mbf F_{k-1}$ and $\mbf J_{k,j}$ will be
\begin{align}
	\mbf F_{k-1} &= \mbf I_3, \\
\begin{split}
	\mbf J_{k,j} &= \lim_{\boldsymbol{\epsilon}\rightarrow\mbf 0} \frac{\Log(\Exp(\boldsymbol{\epsilon})\mbf x_{k,j}^+ {\mbf x_k^-}\tran) - \Log(\mbf x_{k,j}^+ {\mbf x_k^-}\tran)}{\boldsymbol{\epsilon}} \\
	&=\mbf J_l^{-1}(\delta \phi),
\end{split} \\
	\delta \phi_j &= \mbf x_{k, j}^+ \boxminus \mbf x_k^-.
\end{align}

\section{Related Work}
The IEKF for the vector space had been discussed in \eg, \cite{havlik_performance_2015}.
Later, it was extended to the state vector on differentiable manifolds, \eg, \cite{bloesch_iterated_2017}.
However, the definition of the boxplus operator Jacobian $\mbf L$ \cite[(49)]{bloesch_iterated_2017} is vague, 
and $\mbf L$ was erroneously transposed in \cite[(50-52)]{bloesch_iterated_2017}.
In fact, $\mbf L$ relates the multiplicative increment on the manifold to the additive increment in the tangent space (see also \eqref{eq:right-jac}), \ie,
\begin{equation}
\mbf L(\phi) = \lim_{\boldsymbol{\epsilon}\rightarrow\mbf 0}
\frac{\Exp(\phi + \boldsymbol{\epsilon})\boxminus \Exp(\phi)}{\boldsymbol{\epsilon}}
\end{equation}
where $\Exp$ is the exponential map \cite{manfredo_p_do_carmo_differential_2016} for the differentiable manifold.
For vectors, $\Exp(\mbf x) = \mbf x$. For SO3 elements, $\Exp(\phi) = \mbf I_3 + \sum_{k=1}^{\infty}\frac{1}{k!} \crossmat{\phi}^k$.
Authors of FAST-LIO \cite{he_kalman_2021} also formulate the IEKF on manifolds, but at great length with many new notations.
The lack of a quick-start reference for the IEKF on manifolds motivates this note.
Note that X. Gao's recent book ``SLAM technology in autonomous driving and robotics'' also presents a simplified IEKF for lidar-inertial odometry in chapter 8.3.

\appendix
Here are a few helpful equations for the above derivations.
\begin{equation}
\frac{\partial (\mbf a + \mbf{Bx})\tran \mbf P^{-1} (\mbf a + \mbf{Bx})}{\partial \mbf x}
= 2	(\mbf{a + Bx})\tran \mbf P^{-1} \mbf B
\end{equation}
Handy Jacobians for SO3 elements are copied from \cite{hartley_contact-aided_2019},
\begin{align}
\Exp(\phi + \delta \phi) \approx \Exp(\phi)\Exp(\mbf J_r(\phi) \delta \phi) \label{eq:right-jac}\\
\mbf{J}_r(\phi) = \mbf I_3 - \frac{1- \cos \theta }{\theta^2} \crossmat{\phi} + \frac{\theta - \sin{\theta}}{\theta^3} \crossmat{\phi}^2 \\
\Exp(\phi + \delta \phi) \approx \Exp(\mbf J_l(\phi) \delta \phi)\Exp(\phi) \\
\mbf J_r^{-1}(\phi) = \mbf I_3 + \frac{1}{2} \crossmat{\phi} + \left(\frac{1}{\theta^2} - \frac{1 + \cos \theta }{2\theta \sin\theta}\right) \crossmat{\phi}^2 \\
\mbf{J}_r\tran(\phi) = \mbf J_l(\phi) = \mbf J_r(-\phi) \\
\mbf J_r\minustran(\phi) = \mbf J_l^{-1}(\phi) = \mbf J_r^{-1}(-\phi)
\end{align}

\bibliographystyle{IEEEtran}
\bibliography{zotero}

\end{document}